\documentclass[aip,pop,groupedaddress]{revtex4-1}

\usepackage{graphicx}

\begin{document}

\title{Effect of the reference electrode size on the ionization instability in the plasma sheath of
a small positively biased electrode}

\author{Y.P. Bliokh}
\affiliation{Department of Physics, Technion, 32000 Haifa, Israel}
\author{Yu.L. Brodsky}
\affiliation{Department of Physics, Technion, 32000 Haifa, Israel}
\author{Kh.B. Chashka}
\affiliation{Department of Physics, Technion, 32000 Haifa, Israel}
\author{J. Felsteiner}
\affiliation{Department of Physics, Technion, 32000 Haifa, Israel}
\author{Ya.Z. Slutsker}
\affiliation{Department of Physics, Technion, 32000 Haifa, Israel}

\begin{abstract}
It is well known that  additional ionization in the vicinity of a
positively biased electrode immersed into a weakly ionized plasma
is responsible for a hysteresis in the electrode current-voltage
characteristics and the current self-oscillations rise.  Here we
show both experimentally and theoretically that under certain
conditions these phenomena cannot be correctly interpreted once
considered separately from the reference electrode current-voltage
characteristics. It is shown that small electrodes can be
separated into three groups according to the relation between the
electrode and the reference electrode areas. Each group is
characterized by its own dependence of the collected current on
the bias voltage.

\end{abstract}


\maketitle

\section{Introduction}

Phenomena related with a biased electrode (Langmuir probe)
imbedded into a plasma have been studied beginning from the rise
of plasma physics and are still in progress. The aim of the
current investigation is the determination of the link between the
measured current-voltage (C-V) electrode characteristics and the
plasma parameters. For this purpose it is desirable to minimize
the disturbances of the actual plasma parameters introduced by the
probe.

However, there is a number of processes in the surrounding plasma
caused by the probe presence that affect significantly the probe
C-V characteristics. The probe C-V characteristics can be
multi-valued and demonstrate a hysteresis-like behavior. Moreover,
the current collected by the probe produces a set of instabilities
(sheath-plasma instability \cite{Stenzel1}, ionization induced
instabilities \cite{Klinger, Stenzel2}, \textit{etc}). As a
result, the self-oscillations of the collected current rise even
when the probe potential is kept constant.

There are several papers where the physical processes responsible
for the excitation of the probe current oscillations are studied.
However, the role of a mandatory element -- the reference
electrode for the probe -- that closes the current circuit through
the plasma, has to our knowledge not been adequately explored. In
this paper the results of a detailed  experimental investigation
of the C-V characteristics of a positively charged probe are
presented. Various probes have been studied whose area $S_{pr}$ is
small enough  compared with the ion-collected electrode area (the
reference electrode area) $S_i$:
\begin{equation}\label{eq00}
S_{pr}\ll (m_e/m_i)^{1/2}S_i,
\end{equation}
where $m_e$ and $m_i$ are the electron and ion masses,
respectively. It will be shown that probes, whose areas meet the
condition (\ref{eq00}), can be separated into three groups: large,
intermediate, and small probes. Each group is characterized by its
own regime of interaction with the plasma. Affiliation with one
group or another is determined by the relation between the C-V
characteristics of the probe and the reference electrode which are
connected in series via the plasma.

This paper is organized as follows. In section II, the
experimental setup is described. The experimental results are
presented in section III. A simple qualitative theoretical model
and comparison with the experimental results are presented in
section IV. The paper main results are summarized and discussed in
section V.

\section{Experimental setup}

\begin{figure}[tbh]
\centering \scalebox{0.4}{\includegraphics{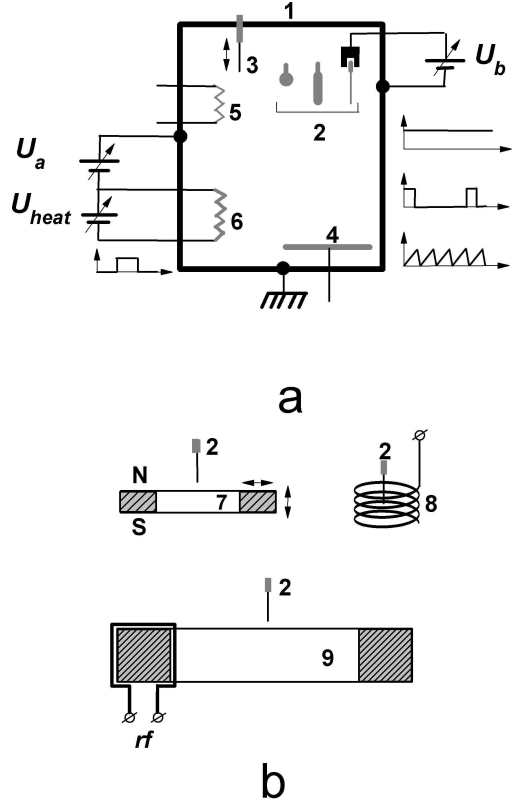}} \caption{
a: 1 -- vacuum chamber; 2 -- set of replaceable probes; 3 --
movable probe; 4 -- large surface probe; 5 -- a thin tungsten
filament (hot probe); 6 -- hot cathode; $U_{heat}$  pulse duration
-- 0.2~s, repetition rate -- 5~s, $U_b$ -- bias voltage, either
dc, or rectangular pulses with 100~$\mu{\rm s}$ duration and
variable delay inside each discharge pulse, or continuous
saw-tooth voltage with 200~$\mu{\rm s}$ period. b: 7 --
replaceable Nd magnet,  inner diameter 5~cm, outer diameter 7.5~cm
and thickness 0.6~cm; 8 --  replaceable spiral around the probe; 9
-  ferromagnetic core with 15 winding turns, inner diameter 10~cm,
outer diameter 15~cm, thickness 2.5~cm. } \label{1}
\end{figure}

The experiments were mainly performed in a stainless steel vacuum
chamber, having an inner diameter of 30~cm and a height of 8~cm
(Fig.~\ref{1}a). This chamber was equipped with two insulated
tungsten filaments, one of them had 0.3~mm diameter and the other
one -- 0.1~mm. Both of them could be heated separately. The first
one could be moved just when the chamber was open and the other
one could be moved inside the chamber during experiments. The
chamber was also equipped with a large surface probe, having an
area $S=7\,\,{\rm cm}^2$, with a  comparatively small ($S\simeq
0.6\,\,{\rm cm}^2$) movable single probe and with a special holder
for replaceable platinum probes of various lengths (1 -- 25)~mm
and diameters (0.05 -- 5)~mm. The first (thicker) filament was
used as a hot cathode. It was heated when a voltage pulse
$U_{heat}$ of 0.2~s duration was applied per each 5~s. To obtain a
hot-cathode discharge and a plasma we applied a dc voltage either
between the cathode and the grounded vacuum chamber or between the
cathode and the large surface probe. In the latter case the
discharge circuit was floating. The plasma density $n_p$ and the
electron temperature $T_e$ were derived from the C-V
characteristics of the above-mentioned probes. To verify these
measurements we  could also use the movable  probe as a resonance
probe \cite{r1}. To measure the plasma potential, the thin
filament was used as a hot probe. To heat it, another dc source
was used. To bias the small platinum probe we used either a dc
power supply or a pulsed saw-tooth or  rectangular voltage with
pulse duration (150 -- 200) $\mu$s (see Fig.~\ref{1}a). Just a
positive bias was used. The pulsed regime was used either to
obtain its C-V characteristics in one ``shot'' or to prevent probe
overheating if the collected current was too high. A thin glass
plate of 3~cm diameter (not shown in Fig.~\ref{1}a) could be
placed between the hot cathode and the small positively biased
probe to prevent direct current of the emitted electrons to this
probe. A strong permanent magnet could be placed in various
positions in the vicinity of the small probe in such a way that a
magnetic field of (100 -- 600) Gauss could be achieved there (see
Fig.~\ref{1}b). Also a set of spirals having a diameter of (5 --
20)~mm and a 25~mm length could be placed around this probe
(Fig.~\ref{1}b). These spirals  could also be biased either
positively or negatively with respect to the probe. Some verifying
measurements with just a hot cathode and a small probe could be
carried out in a big vacuum chamber having a 66~cm diameter and a
100~cm length.

\begin{figure}[tbh]
\centering \scalebox{0.4}{\includegraphics{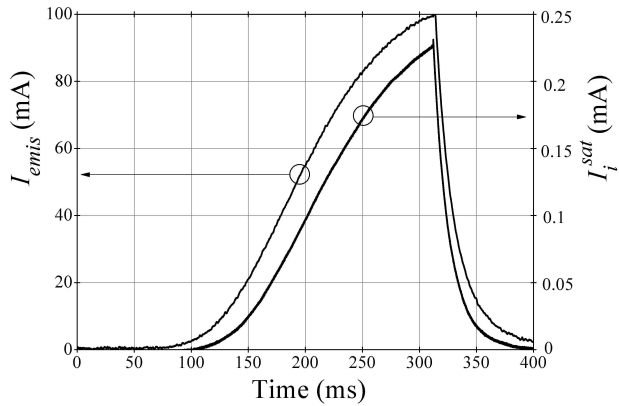}}
\caption{(a)
 cathode emitted current $I_{emis}$, $U_{heat}=10.5$~V,
$U_a=100$~V, Xe gas. (b) ion saturation current $I_i^{sat}$,
collected by the surface probe ($S=7\,{\rm cm}^2$) and multiplied
by factor $4\cdot 10^2$.} \label{2}
\end{figure}

In our experiments we used either Ar at a pressure (0.5 --
5)~mTorr or Xe at a pressure (0.1 -- 1)~mTorr. Typically we
applied an anode voltage   $U_a\sim 100$~V in order to make sure
that we work in a regime with the thermal limitation of the
emitted current. This is reasonable in order to eliminate
dependence of the plasma density $n_p$ on $U_a$. Indeed in the
range $U_a=(80\,-\,100)$~V, variations of the emitted current
$I_{emis}$ did not exceed 10\%. On the other hand it was certain
that in measurements with a large positive bias $U_b$ on the small
probe the value of $U_a$ could be reduced to zero (the cathode
directly connected to the chamber). In the investigated pressure
range the plasma density $n_p$  was found to be directly
proportional to the pressure and the emitted current $I_{emis}$.
The waveform of the emitted current $I_{emis}$ along with the ion
saturation current $I_i^{sat}$, collected by the large surface
probe, are shown in Fig.~\ref{2}. The duration of the $I_{emis}$
growth ($\sim 0.1$~s) was long enough to consider a steady state
plasma at any moment. In our experiments a 100~mA emission current
corresponded to $n_p\leq 3\cdot 10^9\,{\rm cm}^{-3}$, $P=3\cdot
10^{-4}$~mTorr and $T_e\sim(2\,-\,3)$~eV with Xe gas. For the
$n_p$ measurements the discrepancy between the probe
characteristics method and the resonance probe method did not
exceed 15\%. The fraction of fast electrons having energy $\sim
eU_a$  never exceeded (1 -- 2)\%. They appeared as a step in the
ion part of the probe characteristics \cite{r2}. The emitted
current $I_{emis}$  as well as the probe current $I_{pr}$ were
measured with small current-view resistors. In order to study the
influence of the ionization level there was a possibility to
insert inside the vacuum chamber a single-core ferromagnetic
inductively coupled (FIC) plasma  source, driven by a powerful
pulsed rf oscillator \cite{r3} (see Fig.~\ref{1}b). With this
source the plasma density increased a few hundred times (at the
same pressure) and the ionization rate could reach a value of (20
-- 30)\%.

\section{Experimental results}

When the positive bias $U_b$  applied to any of the replaceable
probes (Fig.~\ref{1}a) was relatively low  ($U_b<25$~V  for Xe and
$U_b<40$~V  for Ar) the various C-V characteristics obtained with
our set of probes, along with the measured potential fall
$\Delta\varphi$ between the plasma and the probe, have already
been described and explained long ago (see e.g. Ref. 7 
and references therein). At a higher bias $U_b$ a significant
deviation could appear with various combinations of the gas
pressure $P$ and probe sizes. Note, that at each ``shot'' the
plasma density $n_p$  increases from a very low value, when the
Debye length $r_d$  is large and the plasma sheaths near walls and
electrodes could be comparable to the chamber sizes, to a value
$n_p\geq 5\cdot 10^9\,{\rm cm}^{-3}$, when $r_d$ becomes small
compared to the smallest probes we used.

\begin{figure}[tbh]
\centering \scalebox{0.60}{\includegraphics{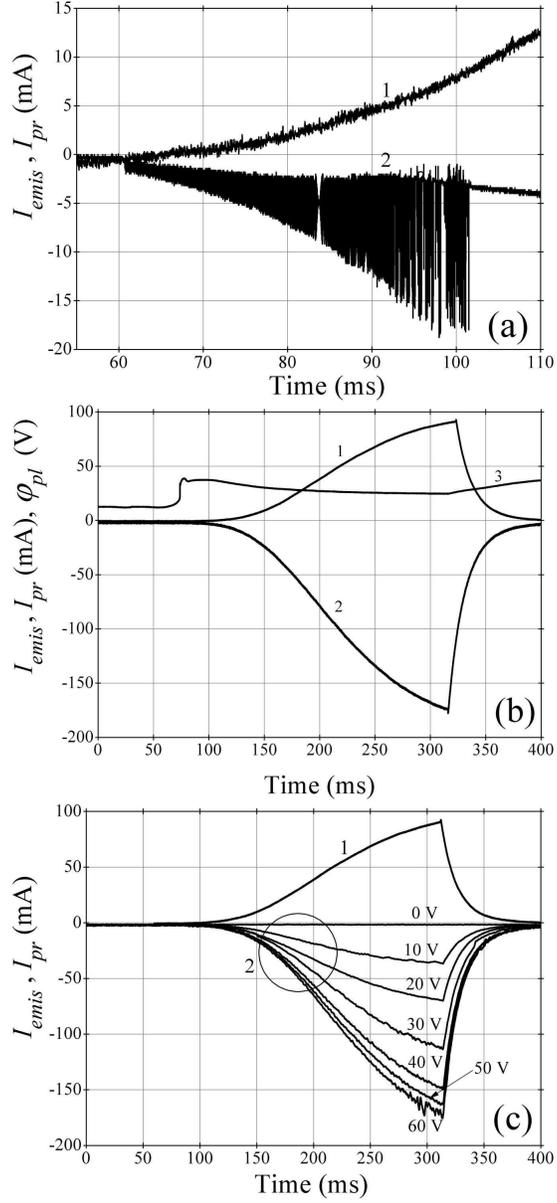}}
\caption{1 -- emitted current $I_{emis}$; 2 -- probe current
$I_{pr}$; $U_a=100$~V, $P=0.3$~mTorr of Xe. (a) beginning of the
discharge pulse, $U_b=40$~V; the $I_{pr}$ irregular perturbations
are clearly seen. (b) 3 -- full waveform of the potential fall
between plasma and walls $\varphi_{pl}$ (the plasma-probe
potential $\varphi_{pr}=U_b-\varphi_{pl}$), waveforms of emitted
current $I_{emis}$ and probe current $I_{pr}$ (the perturbations
in the beginning are filtered out);  $U_b=60$~V. (c) $I_{pr}$
saturation with
 $U_b$ growth.} \label{3}
\end{figure}

\textit{Large probe.} --- Thus, with the largest probe we used
($R=5$~mm) and at a fixed pressure, say $P=0.3$~mTorr of Xe, the
same probe characteristics could be also kept for $U_b>25$~V just
with one exception. Namely at very low plasma densities some
irregular oscillations appeared in the probe current $I_{pr}$ (see
Fig.~\ref{3}a). For higher  $n_p$ the behavior of $I_{pr}$ and the
potential fall $\varphi_{pr}$ between the probe and the plasma
during the pulse are very similar to the previous case when $U_b$
was low (Fig.~\ref{3}b). The maximal electron probe-collected
current $I_{pr}$, saturated at high $U_b$, was equal to the sum of
the emitted current $I_{emis}$ and the total ion saturation
current to the vacuum chamber walls (Fig.~\ref{3}c). The total ion
saturation current to the walls was derived as the geometrical
ratio of the surface probe area ($7\,{\rm cm}^2$) and the total
area of the metal vacuum chamber ($S_a\sim 2\cdot 10^3\,{\rm
cm}^2$) multiplied by the ion saturation current to this probe
$I_i^{sat}$ (Fig.~\ref{2}). In the case of the floating discharge
circuit, i.e. when $U_a$ was applied to the large surface probe
instead of the chamber wall,  $I_{emis}$ was simply subtracted
from this sum. It should be noted that at the densities when the
irregularities of $I_{pr}$ appear, the measurements showed that
the thickness of the plasma sheath near the surfaces become
comparable to the chamber size \cite{r5}. In our case it
corresponds to $n_p\leq 3\cdot 10^8\,{\rm cm}^{-3}$,
$I_{emis}\leq(5\,-\,8)$~mA for Xe, and
$I_{emis}\leq(10\,-\,15)$~mA for Ar. The pressures were 0.3~mTorr
and 1~mTorr for Xe and Ar respectively. Reducing drastically -- in
one order of magnitude -- the pressure $P$ at the same pulse of
$I_{emis}$ (at the same $U_{heat}$) it was possible to obtain
these $I_{pr}$ irregularities during the whole pulse of
$I_{emis}$. Again, if the discharge circuit was floating, the
maximal amplitude of the probe current irregular perturbations
became smaller at the corresponding $I_{emis}$ value. While these
perturbations appeared in the electron probe current $I_{pr}$, no
visible changes were seen neither in the emitted current
$I_{emis}$ (see Fig.~\ref{3}) nor in the ion saturation current
collected by a movable or surface probe. The latter means that the
plasma density $n_p$ kept constant.

\begin{figure}[tbh]
\centering \scalebox{0.60}{\includegraphics{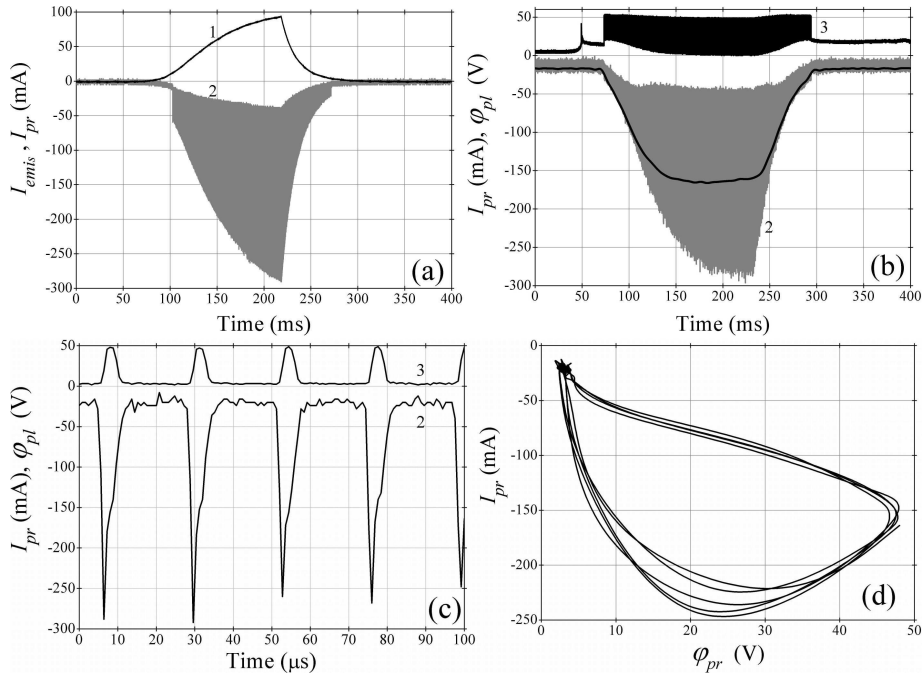}}
\caption{Small probe: 1~mm diameter, 2~mm length, $P=0.3$~mTorr of
Xe, $U_a=100$~V. (a) 1 -- emission current $I_{emis}$; 2 -- probe
current $I_{pr}$; $U_b=58$~V. (b) 3 -- potential fall between the
chamber wall and plasma  $\varphi_{pl}$ (potential fall between
plasma and probe $\varphi_{pr}=U_p-\varphi_{pl}$),  thick line --
average value; $U_b=61$~V. (c) the same, but expanded in the mid
of pulse. (d) probe current $I_{pr}$ vs $\varphi_{pr}$ for the
latter case. } \label{4}
\end{figure}

\textit{Intermediate probe.} --- A very different behavior of
$I_{pr}$ could be observed when the probe sizes were reduced to
(0.6 -- 1.2)~mm. At very low pressures ($P<0.05$~mTorr) the same
irregular oscillations could  still be seen. In the pressure range
(0.1 -- 0.8)~mTorr for Xe, when the probe bias exceeds 25~V, but
still below a certain threshold, the probe C-V characteristics
appeared as a prolongation of those for the lower voltage. They
could be described by the theories mentioned above \cite{r4,
Mott}. On the other hand, when the bias voltage $U_b$ reached a
certain value, a single, few or many current spikes appeared in
the probe current $I_{pr}$. After adding further (2 -- 4)~V,
periodic spikes filled the whole plasma pulse, i.e. in the wide
range of $I_{emis}\propto n_p$ there appeared a probe-current
instability. The corresponding $U_b$  is recognized as the
threshold   $U_{th}$. Namely at $U_{th}=58$~V and at $P=0.3$~mTorr
of Xe in the density range of $n_p\approx(4\cdot 10^8\,-\,4\cdot
10^9)\,{\rm cm}^{-3}$ the periodic spikes in the probe current
$I_{pr}$ are shown in Fig.~\ref{4}a,c. Their period and duration
kept approximately constant in the mentioned above density range
and weakly depended on the bias voltage $U_b>U_{th}$. The waveform
of $I_{pr}$ in combination with the potential fall $\varphi_{pl}$
between the plasma and the wall at the same $I_{emis}$, but at a
bit higher $U_b=61$~V, are shown in Fig.~\ref{4}b. It is
interesting to note, that these periodic spike oscillations, as a
rule, started from a small but finite value of $I_{emis}$, the
stage with irregular oscillations was usually  skipped. On the
other hand, in the narrow and unstable ranges of the bias $U_b$
and pressure $P$ it was possible to obtain irregular oscillations
at low $n_p$ which were switched to regular ones at higher $n_p$.

These spikes were quite narrow compared to their period (see
Fig.~\ref{4}c). Typically the spike duration was
$\sim(1\,-\,1.5)\,\mu{\rm s}$ for Ar and $\sim(2\,-\,3)\,\mu{\rm
s}$ for Xe while the spike period was $\sim(10\,-\,15)\,\mu{\rm
s}$ and $\sim(20\,-\,30)\,\mu{\rm s}$, respectively. It should be
noted that, as  seen in Fig.~\ref{4}c, the spikes of $I_{pr}$
correspond to the minimal potential fall between the plasma and
the probe: $\varphi_{pr}=U_b-\varphi_{pl}$, i.e. $I_{pr}$ and
$\varphi_{pr}$ are in opposite phase. Also, as seen in
Fig.~\ref{4}d, there is a hysteresis in the probe C-V
characteristics when this instability exists.

\begin{figure}[tbh]
\centering \scalebox{0.4}{\includegraphics{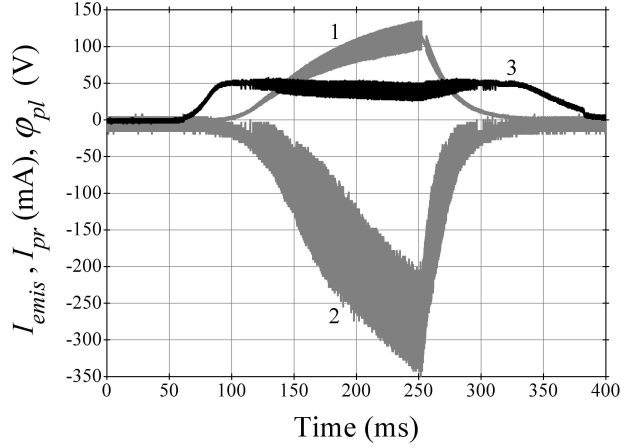}} \caption{
Small probe: 1~mm diameter, 2~mm length, $P=0.3$~mTorr of Xe,
$U_a=100$~V, $U_b=58$~V.  1 -- emission current  $I_{emis}$; 2 --
probe current $I_{pr}$;  3 -- potential fall between the chamber
wall and plasma $\varphi_{pl}$. }  \label{5}
\end{figure}

The threshold bias $U_{th}$  definitely decreased with the
increase of the pressure $P$: for Xe $U_{th}\approx 30$~V  at
$P=1$~mTorr, $U_{th}\approx 58$~V at $P=0.3$~mTorr and
$U_{th}\approx 80$~V at $P=0.1$~mTorr. A very similar tendency was
obtained with Ar. At fixed pressure $P$  the spikes amplitude
increased monotonically with $U_b$  and eventually reached its
maximum which is  equal to the total ion saturation current at the
chamber wall. The average value of  $I_{pr}$  usually did not
exceed 30\% of the spike amplitude (see Fig.~\ref{4}b). The
minimal value of $I_{pr}$ (see Fig.~\ref{4}c) usually corresponded
to the undisturbed (with no spikes) probe current, mentioned above
\cite{r4, Mott}. A further increase of the probe bias $U_b$ led to
qualitative changes of the probe current waveform (see
Fig.~\ref{5}): the spikes minima increase, the spikes amplitudes
decrease, and a visible modulation appears in the ion saturation
current $I_i^{sat}$ collected by the surface probe (Fig.~\ref{5}),
i.e. there appears a  modulation of the plasma density in the
whole plasma volume.

The above probe current instability could be ``killed'' by an
external magnetic field of (600 -- 800)~Gauss. Such a magnetic
field was obtained with a permanent magnet (Fig.~\ref{1}b), placed
in the probe vicinity. In this case the ion gyrofrequency crudely
corresponded to the oscillation frequency with no magnetic field:
35~kHz vs (30 -- 40)~kHz (Xe). Another way to stop this
instability was to put a thin spiral around the probe
(Fig.~\ref{1}b). When the spiral radius was less than (5 -- 7)~mm,
it surely stopped the instability independently whether the spiral
was floated, grounded or biased up to $\pm 30$~V. If the spiral
radius exceeded (10 -- 12)~mm, no  influence was seen. It means
that the processes, determining the instability, were concentrated
in the probe vicinity. Note, that the ion gyroradius $\sim 1$~cm
corresponds to the ion energy  $\sim 5$~eV, which is about
$(0.2\,-\,0.3)\varphi_{pr}$, where $\varphi_{pr}$ is the potential
fall between the probe and the plasma corresponding to the moment
when  $I_{pr}$ is maximal. On the other hand, the existence of
this instability was definitely restricted by the ion current
collected by the opposite, ion collecting electrode, i.e. by the
external circuit. In the case of a small ion collecting area a big
fraction of the probe bias $U_b$ is applied to the plasma sheath
near this electrode. Indeed when this electrode was too small, no
instability appeared. This was directly confirmed when  the ion
collecting electrode we used was either the large surface probe
($S=7\,{\rm cm}^2$) or the insulated metal chamber cap
($S=200\,{\rm cm}^2$). In this case the hot-cathode discharge
circuit (i.e. the plasma creating circuit) and the probe current
circuit were separated and the probe circuit was floating.

\begin{figure}[tbh]
\centering \scalebox{0.4}{\includegraphics{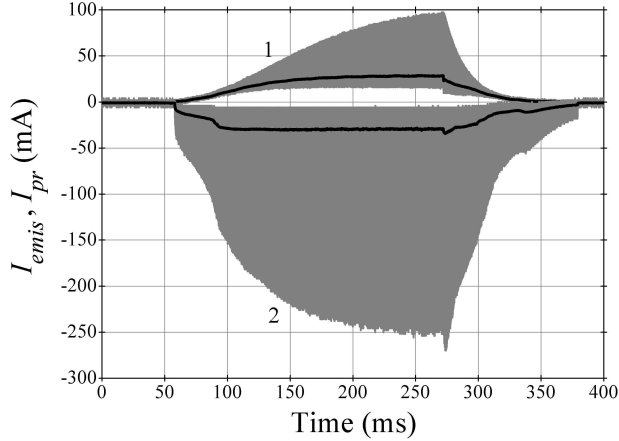}} \caption{
Large vacuum chamber (66~cm diameter, 100~cm length); small probe
(1~mm diameter, 2~mm length); $P=0.3$~mTorr of  Xe,   the hot
cathode is directly connected to the chamber; $U_a=100$~V; 1 --
emission current $I_{emis}$; 2 -- probe current $I_{pr}$. Thick
lines -- corresponding average values.} \label{6}
\end{figure}

It is interesting to note that a similar instability may appear
not only in four-  or three-electrode system, but also in
two-electrode system. Thus, the same hot cathode and the same
probe were placed in the large vacuum chamber (66~cm diameter and
100~cm length). The cathode was directly connected to the chamber
(actually $U_a=0$) and the probe was biased positively. This
scheme is very similar to the one used more than two decades ago
in experiments described by Stenzel\cite{Stenzel1, r6} where, we
believe, similar probe-current spikes were noticed and later were
recognized and studied \cite{Stenzel2}. In our two-electrode
experiments, when $U_b$ exceeded a certain threshold $U_{th}$, the
periodic spikes appeared not only in the probe current but also in
the emitted current $I_{emis}$ (Fig.~\ref{6}). This is contrary to
the three- or four-electrode system, where $I_{emis}$ kept
constant. To obtain these oscillations the required $U_b$ was
approximately the same as in the case described in Refs. 1,10
and it was higher than in the case of three-
or four- electrode system. Again, if the chamber (which is a big
ion collector) was disconnected from the hot filament (cathode),
no oscillation appeared in the probe circuit.

\begin{figure}[tbh]
\centering \scalebox{0.5}{\includegraphics{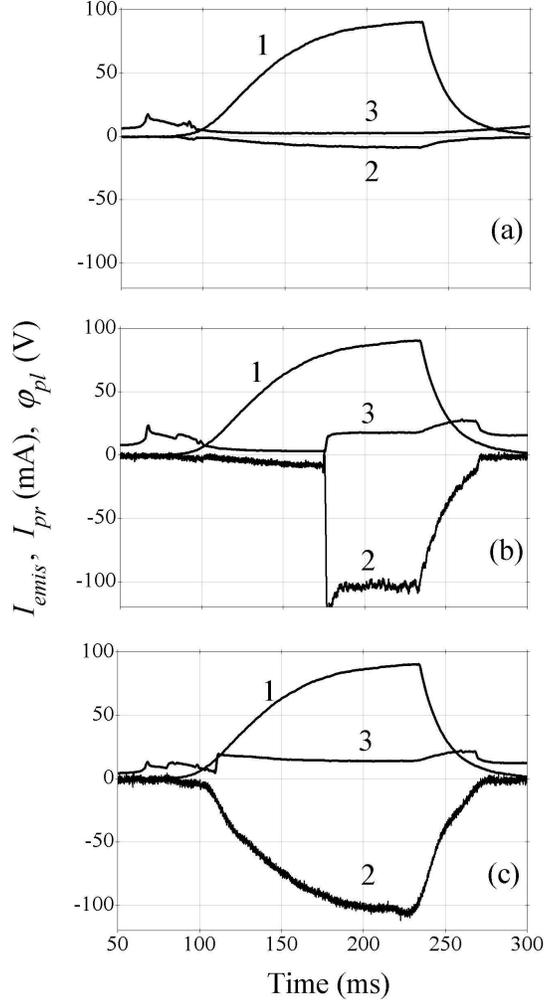}} \caption{
Very small probe: $50\,\mu$  diameter, 1~mm length, $P=0.3$~mTorr
of Xe, $U_a=100$~V. (a) 1 -- emitted current $I_{emis}$; 2 --
probe current $I_{pr}$; 3 -- potential fall between plasma and
wall $\varphi_{pl}$; $U_b=59$~V. (b) the same, but $U_b=61$~V. (c)
the same, but $U_b=62$~V.} \label{7}
\end{figure}

\textit{Small probe.} --- Further decreasing of the probe sizes
again led to a drastic change of the probe characteristics. Thus,
when the probe diameter was below 0.1~mm (with the same length of
2~mm), we could not find any self-consistent oscillations of the
probe current $I_{pr}$. This was correct in the whole investigated
ranges of pressure $P$  and probe bias $U_b$  for both sort of
gases we used. On the other hand, above a certain bias voltage
$U_b$   the probe current $I_{pr}$ jumped from its low value to a
higher one. It happened almost independently of $I_{emis}\sim n_p$
(see Fig.~\ref{7}a). The jump amplitude exceeds one order of
magnitude. The threshold bias voltage $U_{th}$, when the jump
occurs, was close to the threshold bias $U_{th}$, corresponding to
the probe current instability occurrence in the described above
experiments with the larger probe. Obviously, this was correct
just for the same pressure $P$, sort of gas etc.

With further increasing $U_b$ well above $U_{th}$, the probe
current $I_{pr}$ tended to  saturate in the manner shown in
Fig.~\ref{3}c. Actually it also was limited by the total ion
saturation current to the chamber wall. The threshold voltage
$U_{th}$ also decreased with the gas pressure rising up: at
$P=0.3$~mTorr of Xe, $U_{th}\approx 62$~V, and at $P=0.6$~mTorr,
$U_{th}\approx 48$~V. For Ar it was found that $U_{th}\approx
100$~V at $P=1$~mTorr and $U_{th}\approx 70$~V at $P=2$~mTorr.
Unfortunately it was impossible to carry out these measurements in
a wide range of pressures because  a high $U_b$ at low $P$ as well
as a high $P$ at modest $U_b$ could cause parasite sparking which,
in turn, immediately led to the probe evaporation. The
``switching'' time of the probe current jumps was $\sim
20\,\mu{\rm s}$ for Xe (Fig.~\ref{7}b) and $\sim 7\,\mu{\rm s}$
for Ar. These values exceeded by (3 -- 4) times the spike duration
when the probe current instability occurred.

\begin{figure}[tbh]
\centering \scalebox{0.5}{\includegraphics{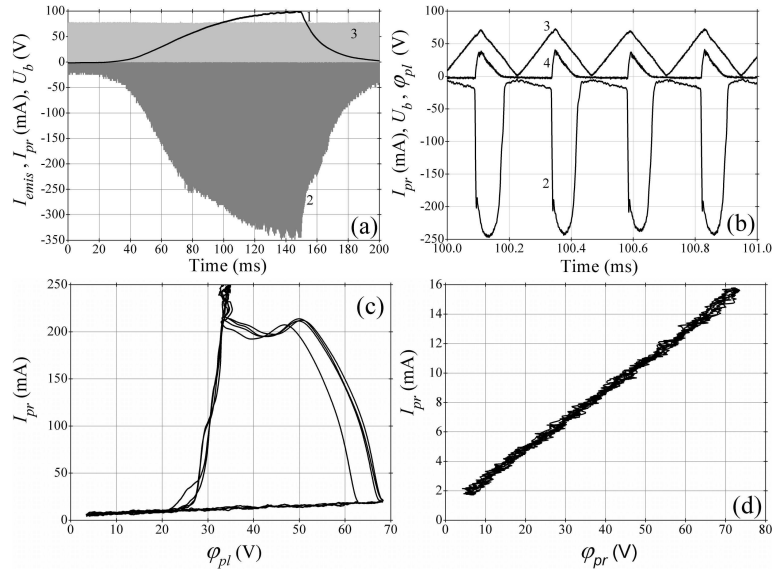}}
\caption{Very small probe: $50\,\mu$  diameter, 1~mm length,
$U_a=100$~V, $U_b$ -- periodic saw-tooth voltage of $200\,\mu{\rm
s}$ period. (a) 1 -- emitted current $I_{emis}$; 2 -- probe
current $I_{pr}$; 3 -- bias voltage  $U_b$; $P=0.3$~mTorr of Xe.
(b) the same as above, but expanded around $t=100$~ms; 4 --
potential fall between plasma and wall $\varphi_{pl}$,
$P=0.3$~mTorr of Xe. (c) hysteresis of the probe characteristics;
$\varphi_{pr}$ -- potential fall between plasma and probe:
$\varphi_{pr}=U_b-\varphi_{pl}$; $P=0.3$~mTorr of Xe. (d) the
same, but $P=0.25$~mTorr  of Xe.} \label{8}
\end{figure}

To study the C-V characteristics of such a thin probe we applied a
periodic saw-tooth voltage instead of the dc $U_b$ (see
Fig.~\ref{8}). The saw-tooth duration was chosen as $\sim
200\,\mu{\rm s}$. On the one hand this is much longer compared to
the switching time, on the other hand this is much shorter
compared to the plasma pulse. The latter means that during
$200\,\mu{\rm s}$ the plasma may be recognized as  stationary. As
it is clearly seen in Fig.~\ref{8}b, the current jump starts at a
substantially higher bias $U_b$ compared to the bias $U_b$ when
the current falls down. It indicates the existence of hysteresis
in the probe C-V characteristics (see Fig.~\ref{8}c). A very
similar hysteresis loop could be seen in a wide range of plasma
densities during the plasma pulse. When the pressure $P$ was a bit
reduced  in order to bring this bias amplitude below $U_{th}$, no
hysteresis was seen and the probe characteristics appeared as an
almost straight line (see Fig.~\ref{8}d), which is in a good
agreement with the known theories for small probes \cite{r4,
Mott}.

\begin{figure}[tbh]
\centering \scalebox{0.4}{\includegraphics{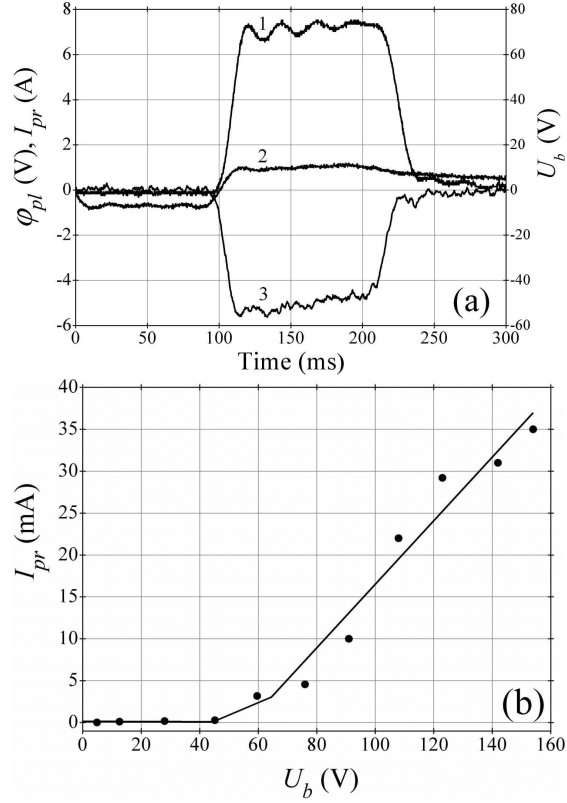}}
\caption{Plasma density  $n_p\simeq 3\cdot10^{12}\,{\rm cm}^{-3}$,
$P=0.3$~mTorr of Xe, ionization rate $\sim 30$\%; very small
probe: $50\,\mu$ diameter, 1~mm length. (a) 1 --  bias voltage
pulse $U_b$; 2 -- potential fall between plasma and wall
$\varphi_{pl}$; 3 -- probe current $I_{pr}$. (b) electron current
collected by the probe vs $U_b$.} \label{9}
\end{figure}

To complete this study we checked the qualitative influence of the
ionization level on the probe characteristics. Namely we placed in
the vacuum chamber the powerful pulsed single-core FIC plasma
source \cite{r3} (see Fig.~\ref{1}b). It could work in the same
pressure range and in our experiments we achieved ionization level
$\leq 30$\%. The plasma density grew up to $\sim 3\cdot
10^{12}\,{\rm cm}^{-3}$ during a $2\,{\rm ms}$ pulse. To prevent
the probe damage, we biased the probe with a single variable
positive voltage pulse with  pulse duration of $\sim 100\,\mu{\rm
s}\ll 2\,{\rm ms}$. Typical waveforms of the probe current and
plasma potential are shown in Fig.~\ref{9}. In the investigated
range of $U_b$ the electron probe current could reach a few
Amperes but the probe characteristics were smooth with no
hysteresis, no instability of the probe current was seen and the
potential fall $\varphi_{pl}$ between the plasma and wall was
always low.

\section{Theoretical model and analysis of the experimental data}

The experimental data show that the peculiarities of the probe C-V
characteristics (instability, hysteresis) are the same either in
three or four electrodes schemes. It means that the discharge
circuit can be eliminated from the consideration.

A positively biased probe collects electron current $I_{pr}$ from
the surrounding plasma. The equivalent ion current $I_{ref}$ is
collected by a reference electrode (the chamber wall in the
experiments). The current loops through an external circuit which
maintains the voltage $U_b$ between the electrodes. The probe and
the reference electrode (thereafter cathode) currents depend on
the voltage between them and the plasma:
$I_{pr}=I_{pr}(\varphi_{pr}-\varphi_{pl})$ and
$I_{ref}=I_{ref}(\varphi_{pl}-\varphi_{ref})$, where
$\varphi_{pr}$, $\varphi_{pl}$, and $\varphi_{ref}$ are the probe,
plasma, and cathode potentials, respectively (thereafter
$\varphi_{ref}=0)$. Thus, taking into account that
$\varphi_{pl}+\varphi_{pr}=U_b$, the plasma potential
$\varphi_{pl}$ is determined by the condition:
\begin{equation}\label{eq0}
I_{pr}(U_b-\varphi_{pl})=I_{ref}(\varphi_{pl}).
\end{equation}
In other words, the plasma potential and, consequently, the probe
current are determined by  both the probe and cathode C-V
characteristics.

It is clear now that the dependence $I_{pr}(U_b)$, which we
measure in our experiments, is close to the probe characteristics
$I_{pr}(\varphi_{pl})$ when the plasma potential is small,
$\varphi_{pl}\ll U_b$. This is possible when the ion-collecting
area $S_i$ (cathode area) is large enough  compared with the probe
area $S_{pr}$ and Eq.~(\ref{eq00}) is satisfied.

When the bias voltage is large, $eU_b\gg T_e$, the effective
collecting area $S_{eff}$ around the small probe can exceed the
probe geometric area and the condition in Eq.~(\ref{eq00}) should
be written as follows:
\begin{equation}\label{eq2}
S_i\gg (m_i/m_e)^{1/2}S_{eff}.
\end{equation}

When the voltage $U_b$ exceeds the ionization potential of the
neutral gas, the ionization in the space-charge sheath around the
probe can change drastically the collected current even if the
electron impact ionization mean free path is large  compared with
the sheath thickness (see, e.g., Refs. 11-13
). The probe C-V characteristics becomes three-valued with
unstable middle branch and demonstrates hysteresis under a gradual
increase and a subsequent decrease of the probe potential
\cite{Ionita}. The collected current on the stable upper and lower
branches of the hysteresis loop can differ by more than an order
of magnitude \cite{Klinger, Strat}.

\begin{figure}[tbh]
\centering \scalebox{0.4}{\includegraphics{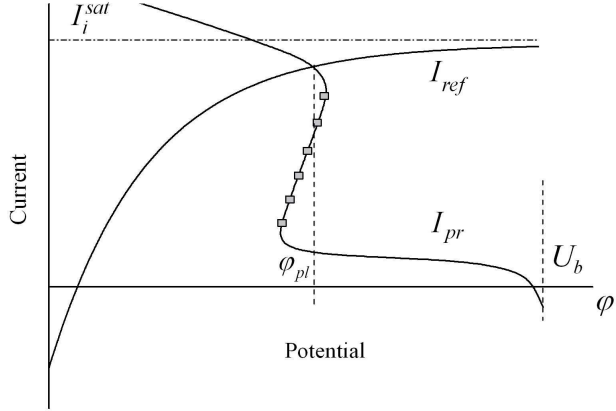}}
\caption{Probe and reference electrode C-V characteristics. Gray
squares mark the unstable branch of the characteristics.}
\label{Fig_1}
\end{figure}

When the probe and the cathode C-V characteristics are known, the
plasma potential $\varphi_{pl}$ that satisfies Eq.~(\ref{eq0}) can
be found graphically as it is depicted in Fig.~\ref{Fig_1}.
The cathode C-V characteristics is shown as a function of the
potential $\varphi$, $I_{ref}(\varphi)$, whereas the probe
characteristics is shown as a function of $U_b-\varphi$,
$I_{pr}(U_b-\varphi)$. The intersection of these two curves
defines the plasma potential $\varphi_{pl}$ and the circuit
current $I(U_b)$. Under the increase/decrease of the bias voltage
$U_b$, the probe C-V characteristics curve is shifted right/left
along the horizontal axis as a whole.

It is important to bear in mind that the probe characteristics
shape depends on the probe dimension and form. Further we will
consider small  probes whose radii are comparable with the plasma
Debye length. Such probes are characterized by the absence of a
saturation current \cite{Mott}. The characteristic values of the
current which are inherent in the $I_{pr}(\varphi)$ dependence
(characteristic scale along the vertical axis in Fig.~\ref{Fig_1})
increase/decrease monotonically with the increase/decrease of the
probe radius.

Now, using this graphical representation, one can explain the
experimental results described above. It is convenient to group
the results in accordance with the probe radius: large,
intermediate, and small probes.

\begin{figure}[tbh]
\centering \scalebox{0.4}{\includegraphics{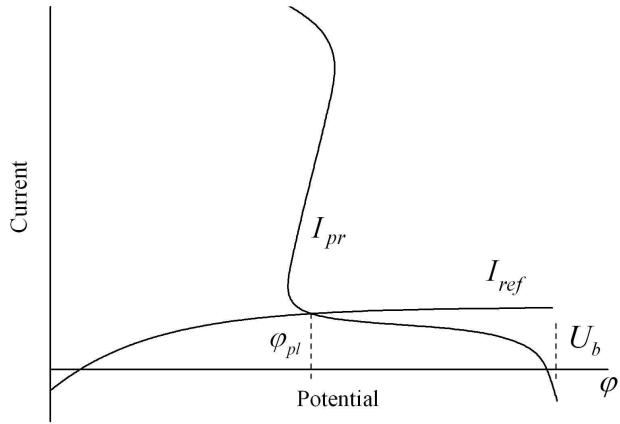}}
\caption{Large probe.} \label{Fig_4}
\end{figure}

\textit{Large probe.} --- When the probe radius is so large that
the collected current exceeds the cathode saturation current even
on the lower branch of the probe C-V characteristics, the
intersection has the form depicted in Fig.~\ref{Fig_4}.
Independently of the bias voltage, the cathode characteristics
intercepts only the lower \textit{stable} branch of the probe
characteristics.
The experiments with the 5mm radius probe are related to this case
(see Fig.~\ref{3}).

\begin{figure}[tbh]
\centering \scalebox{0.4}{\includegraphics{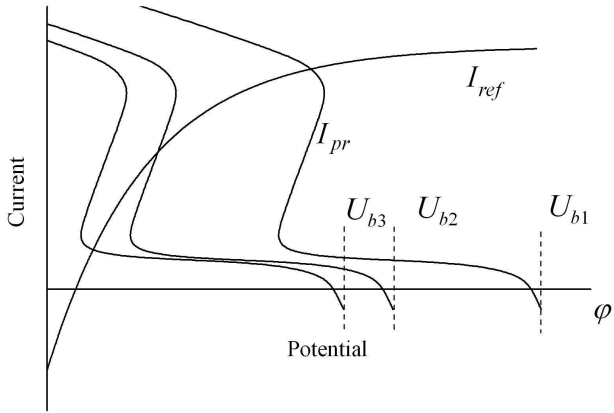}}
\caption{Intermediate probe. $U_{b1}<U_{b2}<U_{b3}$.}
\label{Fig_3}
\end{figure}

\textit{Intermediate probe.} --- Under the probe radius decrease
its C-V characteristics  ``shrinks'' along the vertical axis. The
possible types of the intersections between the probe and cathode
characteristics are shown in Fig.~\ref{Fig_3}.

The distinctive property of the intermediate probe is the
following: there is a range of  bias voltage values when the
cathode characteristics $I_{ref}(\varphi)$ intercepts only the
\textit{unstable branch} of the probe characteristics
$I_{pr}(U_b-\varphi)$ (curve $I_{pr}^{(2)}$, bias voltage $U_{b2}$
in Fig.~\ref{Fig_3}). It means that there are no stable solutions
of Eq.~(\ref{eq0}) and self-oscillations of the current in the
circuit appear when the bias voltage lies in the above-mentioned
range. Note that jumps from the lower to the upper branch and
backwards occur in different points of the probe C-V
characteristics. Therefore, the dependence $I_{pr}[\varphi(t)]$
should  demonstrate the hysteresis-like behavior.

The experiments with 0.6-1.2~mm radius probes demonstrate these
peculiarities of the intermediate probe characteristics (see
Fig.~\ref{4}).

\begin{figure}[tbh]
\centering \scalebox{0.4}{\includegraphics{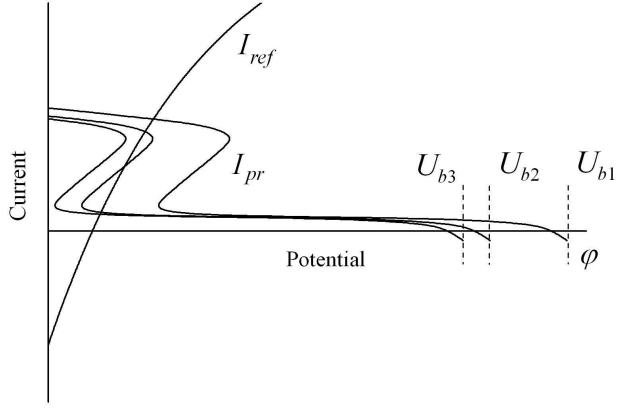}}
\caption{Small probe. $U_{b1}<U_{b2}<U_{b3}$.} \label{Fig_2}
\end{figure}

\textit{Small probe.} ---  If the probe radius is so  small that
the collecting current is small  compared with the ion saturation
current for all voltages in the range of interest, the
intersection between the C-V curves in Fig.~\ref{Fig_1} is located
near the cathode floating potential  (potential that corresponds
to
 zero current). The intersection types for various bias
voltages $U_b$ are shown in Fig.~\ref{Fig_2}.

Depending on the bias voltage, the cathode C-V characteristics
intercepts either the lower branch of the probe characteristics
(curve $I_{pr}^{(1)}$, bias voltage $U_{b1}$ in Fig.~\ref{Fig_2}),
or three branches (curve $I_{pr}^{(2)}$, bias voltage $U_{b2}$),
or the upper branch (curve $I_{pr}^{(3)}$, bias voltage $U_{b3}$).
It is important to note that independently of the bias voltage the
cathode characteristics always intercepts the \textit{stable
branches} (upper and lower branches, the intermediate branch is
always unstable) of the probe characteristics. In this case there
are stable solutions (either one or two) of Eq.~(\ref{eq0}) for
all values of the bias potential. The C-V characteristics $I(U_b)$
demonstrates hysteresis under a gradual increase and a subsequent
decrease of the potential $U_b$.

\begin{figure}[tbh]
\centering \scalebox{0.8}{\includegraphics{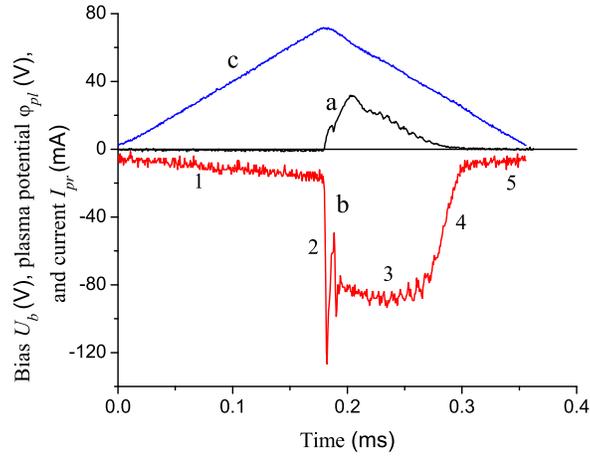}}
\caption{Plasma potential $\varphi_{pl}$ (b) and current $I_{pr}$
(c) variations under a gradual increase and a subsequent decrease
of the bias voltage $U_b$ (a). The numerals that indicate the
consecutive stages of the current variation are used below in
Figs. \ref{Cathode_VA} and \ref{Probe_VA}.} \label{Period}
\end{figure}

\begin{figure}[tbh]
\centering \scalebox{0.8}{\includegraphics{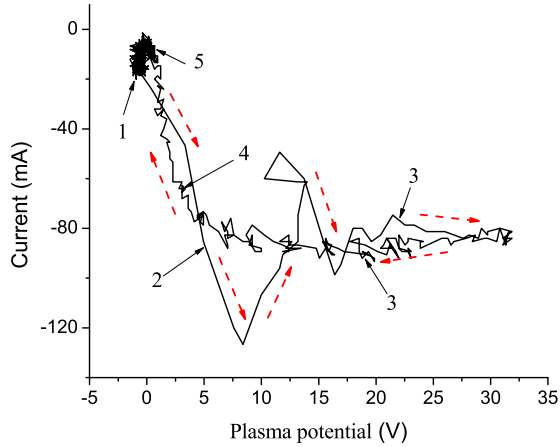}}
\caption{The dependence $I_{ref}(\varphi_{pl})$ under a slow
gradual increase and a subsequent decrease of the potential $U_b$.
The numerals establish correspondence with the consecutive
variation of the probe current shown in Fig.~\ref{Period}. The red
dashed arrows show motion of the representative point in time.}
\label{Cathode_VA}
\end{figure}

\begin{figure}[tbh]
\centering \scalebox{0.8}{\includegraphics{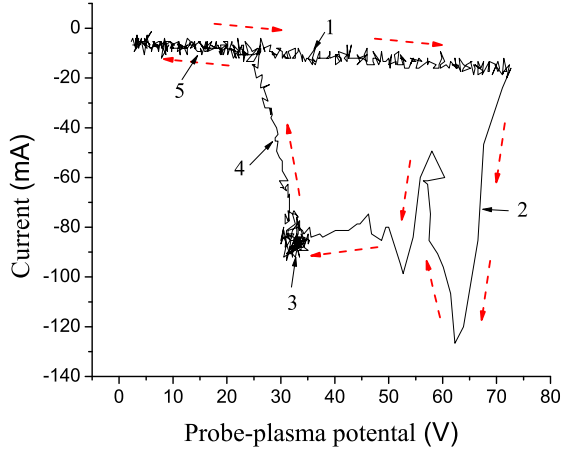}}
\caption{The dependence $I_{pr}(U_b-\varphi_{pl})$  under a slow
gradual increase and a subsequent decrease of the potential $U_b$.
The meaning of the numerals and the red dashed arrows is the same
as in Fig.~\ref{Cathode_VA}} \label{Probe_VA}
\end{figure}

This qualitative analysis explains the experimental results
related to the small (less than 0.1 mm diameter) probe  presented
in Figs.~\ref{Period}, \ref{Cathode_VA}, and \ref{Probe_VA}.

The current circuit contains two elements: the plasma-probe sheath
and the plasma-cathode sheath. The first one is characterized by a
multi-valued C-V characteristics $I_{pr}(U_b-\varphi)$, whereas
the C-V characteristics of the second one is a single-valued
function $I_{ref}(\varphi)$. It means that under a gradual
increase and a subsequent decrease of the potential $U_b$, the
dependence $I_{ref}(\varphi)$ should be a single-valued function,
despite the fact that the dependence $I_{pr}(U_b-\varphi)$
demonstrates hysteresis. This conclusion is confirmed by the
experimental data shown in Figs.~\ref{Cathode_VA} and
\ref{Probe_VA}. Relatively small deviations from the single-valued
dependence $I_{ref}(\varphi)$ (small hysteresis loops) reflect
transient processes of the plasma-cathode sheath reconstruction
caused by fast variations of the plasma potential $\varphi$ under
the ``jumps'' between the lower and upper branches of the probe
C-V characteristics \cite{r2, Bliokh}.

The key element in the qualitative analysis above is the
additional ionization inside the current-collecting sheath around
the probe. Additional ions can essentially increase the sheath
radius if the number of ions $N_i$ in the sheath is comparable
with the number of electrons $N_e$. Assuming that the ionization
take place in all the sheath volume, it is possible to make the
following rough estimate of  $N_i$.   Any electron entering the
sheath produces $n_0\sigma_i R$ ions, where $n_0$ is the gas
density, $\sigma_i$ is the ionization cross section, and $R$ is
the sheath radius. Thus, $N_i/N_e\sim (n_0\sigma_i R)(t_e/t_i)$,
where $t_{e,i}\simeq R/v_{e,i}$ is the characteristic residence
time of the particle in the sheath, and $v_{e,i}$ is the
characteristic velocity of the particle. Because electrons and
ions are accelerated by the same potential drop,
$v_e/v_i=\sqrt{m_i/m_e}$, where $m_i$ is the ion mass. Thus,
\begin{equation}\label{eq3}
N_i/N_e\sim n_0\sigma_i R\sqrt{m_i/m_e}
\end{equation}
Under typical experimental conditions ($n_0\sim 10^{13}{\rm
cm}^{-3}$, $\sigma_i\sim 4\cdot 10^{-16}{\rm cm}^2$,
$\sqrt{m_i/m_e}\sim 4\cdot 10^2$) the number of ions $N_i$ in the
sheath is comparable with the number of electrons $N_e$,
$N_i/N_e\simeq 1$, when the sheath radius is of the order of 1cm.
This estimation is in a good agreement with the experimental data.
Indeed, approximately the same radius $R\sim1$ cm of the probe
collecting area may be obtained directly from the amplitude of the
probe current spikes (Fig.~\ref{4} -- Fig.~\ref{8}). This is
correct in a wide range of plasma densities [emission currents
$I_{emis}=(10 - 100)$ mA]. Experiments with various spirals placed
around the probe showed also that the spiral did not disturb the
current-collecting area when the spiral radius exceeds 1~cm.

The above estimation of the  number of ions $N_i$ in the sheath
supposes implicitly that the neutral gas density $n_0$ is large
enough so as to provide creation of this amount of ions. However,
it does not always happen. In the vicinity of a small probe the
electron density exceeds many times the plasma density due to the
geometric focusing of the collected particles. The ionization rate
in this region can be so large that the ionization degree
$n_p/n_0$ would reach 100\% during the ion residence time or even
faster. In this case the number of ions  $N_i$ is always smaller
than the number of electrons $N_e$ in the sheath and the probe C-V
characteristics is single-valued.

Let us estimate the sheath region where the ionization degree is
small, $n_p/n_0\ll 1$ and the condition
\begin{equation}\label{eq4}
N_i/N_e\sim n_0\sigma_i R\sqrt{m_i/m_e}\sim 1
\end{equation}
is satisfied. The ions production rate in a layer of radius $r$ is
described by the following equation:
\begin{equation} \label{eq5}
{dn_i(r)\over dt}=n_0\sigma_i n_e(r)v_e(r),
\end{equation}
where $n_e(r)$ and $v_e(r)$ are the collected electrons density
and velocity, respectively. During the ion residence time $t_i\sim
R/v_i$ the ionization degree reaches the value:
\begin{equation}\label{eq6}
{n_i(r)\over n_0}\sim n_0\sigma_i n_e(r)v_e(r){R\over v_i}\sim
n_0\sigma_i R\sqrt{m_i/m_e}{n_0\over n_p}\sqrt{T_e/eU_b}{R^2\over
r^2}.
\end{equation}
Here the particles flux conservation law
$n_e(r)v_e(r)r^2=const=n_p\sqrt{T_e/m_e}R^2$ is used.

When the condition (\ref{eq4}) is satisfied, the ionization degree
is small, and $n_p/n_0\ll 1$ in a sheath region of radius
\begin{equation}\label{eq7}
r\gg R\left({n_p\over n_0}\right)^{1/2}\left({T\over
eU_b}\right)^{1/4}\equiv r_c.
\end{equation}
It follows from Eq.~(\ref{eq7}) that the region, where the neutral
gas density cannot provide production of the required amount of
ions, occupies a considerable part of the sheath volume when the
plasma density is large enough, all other factors being the same.

In the set of experiments where plasma is created by discharge
with a hot cathode $(n_p\simeq 3\cdot 10^9\,{\rm cm}^{-3})$ the
ratio $r_c/R\simeq 10^{-2}$ is small and the probe C-V
characteristics is multi-valued. In contrast, in the experiments
with the FIC plasma source $(n_p\simeq 3\cdot 10^{12}\,{\rm
cm}^{-3})$ this ratio is large, $r_c/R\simeq 0.5$, and hence the
probe current is stable and self-oscillations are not observed
despite that the potential fall between the probe and the plasma
could significantly exceed the ionization potential.

\section{Summary}

The characteristics of small positively biased electrodes (probes)
immersed into a plasma have been studied. These probes were much
smaller than the reference electrode size divided by the square
root of the ion/electron mass ratio. The electron current branch
of the C-V characteristics of such probes is widely used to
measure the local plasma parameters. Additional ionization of
neutral gas by electrons accelerated in the space-charge sheath
around the probe could significantly expand  the sheath thickness.
However, we were focused on the case, when the sheath thickness
still remains small compared to the whole plasma and reference
electrode sizes. It should be emphasized that in our experiments
this additional ionization inside the sheath does not change the
plasma parameters in the whole volume. When the probe dimension is
large compared to the sheath thickness, the sheath expansion does
not affect the current collected by the probe from the plasma. On
the other hand, near a very small probe the sheath expansion
induced by the ionization leads to considerable increase of the
collected current.

The probe current loop is closed through the ion-collecting
electrode (cathode) having its own C-V characteristics. The
potential fall between the cathode and the plasma should rise to
support the increasing current. The plasma potential growth
changes the potential drop between the probe and the plasma, and
the probe current should be changed in accordance with the probe
C-V characteristics. Thus, the current is determined by the
relation between the probe and the reference electrode C-V
characteristics.

It has been shown both experimentally and theoretically that these
small probes can be separated into three groups according to the
probe dimensions. Every group is characterized by its particular
dependence of the collected current on the bias voltage. The
current collected by the ``large'' probes increases monotonically
and continuously with the voltage growth. The probes with an
``intermediate'' size are characterized by excitation of strong
periodic spike-like oscillations under a constant bias voltage.
The current collected by the ``small'' probes is stable, but
changes stepwise under certain bias voltage values and
demonstrates hysteresis under a gradual increase and a subsequent
decrease of the bias voltage.  Both oscillations and current steps
are caused by additional ionization in the probe vicinity. Note,
that the same probe can be treated as either large, intermediate,
or small depending on the value of the reference electrode area.
Ignoring the possibility of probe-current instability, neither the
plasma density nor the electron temperature could be measured
correctly by this probe even if the oscillations are filtered out,
averaged etc. In the case of stepwise jumps of the probe current
the upper current level does not reflect directly the plasma
parameters in the place where the probe is located.

In conclusion, we have demonstrated that under certain conditions,
phenomena related with Langmuir probes cannot be correctly
interpreted once considered separately from the reference
electrode characteristics.

\end{document}